\documentclass[sn-basic]{sn-jnl}
\usepackage{graphicx}%
\usepackage{multirow}%
\usepackage{amsmath,amssymb,amsfonts}%
\usepackage{amsthm}%
\usepackage{mathrsfs}%
\usepackage[title]{appendix}%
\usepackage{xcolor}%
\usepackage{textcomp}%
\usepackage{manyfoot}%
\usepackage{booktabs}%
\usepackage{algorithm}%
\usepackage{algorithmicx}%
\usepackage{algpseudocode}%
\usepackage{hyperref}
\usepackage{listings}%
\usepackage{color,soul}
\hypersetup{colorlinks,linkcolor=blue,citecolor=blue,urlcolor=blue}
\usepackage{comment}

\theoremstyle{thmstyleone}%

\theoremstyle{thmstyletwo}%

\theoremstyle{thmstylethree}%

\begin{document}

\title[The infrared arc around $\alpha$ Crucis: a weak interaction with a structured interstellar medium]{The infrared arc around $\alpha$ Crucis: a weak interaction with a structured interstellar medium}

\author*[1]{\fnm{Naira} \sur{Azatyan}}\email{nayazatyan@bao.sci.am}

\author[1]{\fnm{Elena} \sur{Nikoghosyan}}\email{elena@bao.sci.am}

\author[1]{\fnm{Marine} \sur{Torosyan}}\email{marinetorosyan892@gmail.com}

\author[1]{\fnm{Haik} \sur{Harutyunian}}\email{hhayk@bao.sci.am}
\author[1]{\fnm{Derenik} \sur{Andreasyan}}\email{derenik@bao.sci.am}
\author[1,2]{\fnm{Vardges} \sur{Grigoryan}}\email{vardgesgrigoryan99@gmail.com}
\author[1,2]{\fnm{Arman} \sur{Khachatryan}}\email{xachatryanarman566@gmail.com }
\author[1]{\fnm{Anahit} \sur{Samsonyan}}\email{anahit.sam@gmail.com}

\affil*[1]{\orgname{Byurakan Astrophysical Observatory}, \orgaddress{\city{Byurakan vil.}, \postcode{0213}, \country{Armenia}}}

\affil[2]{\orgname{Yerevan State University}, \orgaddress{\street{1 Alex Manoogian}, \city{Yerevan}, \postcode{0025}, \country{Armenia}}}

\abstract{\begin{abstract}

Infrared arc-shaped structures around early-type stars are commonly interpreted as stellar wind bow shocks, although their morphology may also be shaped by large-scale gas motions and density variations in the surrounding interstellar medium. We present a multiwavelength study of the infrared arc associated with the nearby $\alpha$~Crucis system, comprising HD~108248 (B0.5\,IV), HD~108249 (B1\,V), and HD~108250 (B4\,V). We combine \textit{WISE} mid-infrared imaging, \textit{Herschel} far-infrared imaging with dust-temperature and column-density maps, \textit{Planck}  microwave imaging, stellar astrometry, and \textit{eROSITA} X-ray data to investigate the nature of the structure and its relation to the individual stellar components. The \textit{WISE} W4 ridge extends over $\sim$0.65~pc and has a median PSF-corrected width of $\sim$0.037~pc. The HD~108248/HD~108249 pair lies $\sim$0.13~pc from the ridge, compared with $\sim$0.17~pc for HD~108250. Relative to the mean motion of the Lower Centaurus--Crux (LCC) subgroup of the Scorpius--Centaurus association, the close pair is consistent with the mean LCC motion, whereas HD~108250 has a transverse peculiar velocity of only $\sim$1.8~km\,s$^{-1}$. Its motion lies within $\sim$23$^\circ$ of the geometrical symmetry axis of the infrared arc, but points in the opposite sense from the arc. \textit{Herschel} maps show slightly higher median dust temperature ($\Delta T_{\rm d}\sim$0.8~K) and molecular-hydrogen column density ($\Delta N_{\rm H_2}\sim$7$\times$10$^{20}$~cm$^{-2}$) toward the W4-defined region than in the nearby background. On larger scales, the \textit{Planck} maps reveal a strongly structured microwave environment around the system, showing that the infrared arc is projected against a non-uniform large-scale background. Both the HD~108248/HD~108249 pair and HD~108250 are detected by \textit{eROSITA}, with the former being approximately 60 times more luminous in X-rays. The combined morphology, kinematics, and dust properties do not support an unambiguous classical high-velocity bow-shock interpretation. Instead, the arc likely traces a weak interaction between the multiple stellar system and an inhomogeneous ambient medium.

\end{abstract}}

\keywords{Stars: early-type -- Stars: winds -- Infrared: ISM -- X-rays: stars}

\maketitle

\section{Introduction}
\label{sec:introduction}

Massive stars strongly influence their surroundings through ionizing radiation and stellar winds, injecting energy and momentum into the interstellar medium (ISM). When a star moves supersonically relative to the surrounding gas, the interaction between its stellar wind and the ambient medium can produce a bow shock. Such structures can reach characteristic sizes of order a parsec and are frequently detected as arc-shaped nebulae at infrared wavelengths, where dust swept up and heated by the star emits strongly \citep[e.g.][] {buren88,Noriega1997,peri12,kobulnicky16,Mackey2023, Carretero-Castrillo2025}. The morphology of these arcs, together with the stellar kinematics, has therefore been widely used to identify candidate runaway stars and to probe their interaction with the local ISM \citep[e.g.][]{azatyan26}.

An arc-shaped infrared nebula, however, does not necessarily imply a classical wind bow shock produced by a runaway star. Its morphology depends on the relative motion between the star and the surrounding gas rather than on the stellar velocity alone \citep{vandenEijnden2026}. Bulk flows and density gradients in the ISM can modify both the orientation and shape of the interaction region \citep[e.g.][]{Bodensteiner18}. Moreover, wind bubbles around slowly moving massive stars can also produce bright mid-infrared arcs resembling bow shocks \citep{mackey16}. Consequently, distinguishing between a motion-driven wind bow shock, an asymmetric stellar-wind bubble, and a pre-existing interstellar structure requires both stellar kinematic information and measurements of the physical properties of the surrounding material. Observations on multiple spatial scales are particularly important in such cases: high-resolution mid- and far-infrared images trace the local interaction region, whereas lower-resolution all-sky observations can provide the broader environmental context.

The nearby $\alpha$~Crucis system provides an interesting example of this ambiguity. It comprises the close pair HD~108248 and HD~108249, classified as B0.5\,IV and B1\,V stars, respectively \citep{Houk1975,Reed2005}, and the B4\,V star HD~108250 \citep{Tokovinin1999,Reed2005}, located approximately 90~arcsec from the pair. Throughout this work, we adopt a distance of $106.4\pm1.5$~pc for the system, derived from the \textit{Gaia}\,DR3 \citep{Gaia22} parallax of HD~108250, because reliable \textit{Gaia} astrometry is not available for the bright HD~108248/HD~108249 pair. 

An arc-like structure was previously identified as an infrared nebula associated with the bright HD~108248/HD~108249 pair by \citet{Bodensteiner18}, who classified it among the unaligned offset bow-shock candidates around early-type stars. Its pronounced curvature and proximity to the stellar components naturally suggest a possible stellar-wind interaction with the surrounding ISM, although its orientation and large-scale environment make a simple bow-shock interpretation uncertain.

In a previous study, \citet{Torosyan24} investigated the origin and kinematics of the $\alpha$~Crucis system and showed that its motion is consistent with that of the Lower Centaurus--Crux (LCC) subgroup of the Scorpius--Centaurus association. The system lies near the periphery of the projected distribution of the LCC members identified in that study. It was therefore not classified as a high-velocity runaway, despite the presence of the infrared arc. In addition, \citet{Torosyan24} found that the direction of the system's motion is inconsistent with the symmetry axis of the infrared structure. This result suggested that large-scale motions of the surrounding ISM might contribute to the observed morphology. However, the nature of the infrared structure and the possible roles of the individual stellar components remained uncertain.  

In this work, we revisit the infrared environment of the $\alpha$~Crucis system using a multiwavelength approach. Our aim is to investigate the relation between the infrared arc, the individual stellar components, and the surrounding ISM, and to assess whether the observed structure is consistent with a conventional stellar wind bow shock or instead reflects an interaction with a structured ambient medium and, ultimately, clarify the physical origin of the arc.

The paper is organized as follows. In Section~\ref{sec:data}, we describe the data and analysis methods. The observational results, including the morphology of the arc, stellar kinematics, and the distributions of dust temperature and hydrogen column density, are presented in Section~\ref{sec:results} and discussed in Section~\ref{sec:discussion}. Our main conclusions are summarized in Section~\ref{sec:conclusions}.

\section{Data and analysis}
\label{sec:data}

\subsection{Observational data}
\label{obs}
We investigated the environment of the $\alpha$~Crucis system using archival data. The morphology of the surrounding infrared emission was examined using \textit{WISE} imaging \citep{wright10}, with particular emphasis on the W4 ($22\,\mu$m) band, where a prominent arc-like structure is visible close to the stellar system. The same region was inspected in the \textit{Herschel} Photodetector Array Camera and Spectrometer (PACS; \citealt{Poglitsch2010}) and Spectral and Photometric Imaging Receiver (SPIRE; \citealt{Griffin2010}) bands in order to trace the far-infrared emission and investigate how the morphology of the structure changes toward longer wavelengths.

\begin{figure}
\includegraphics[width=5.5cm]{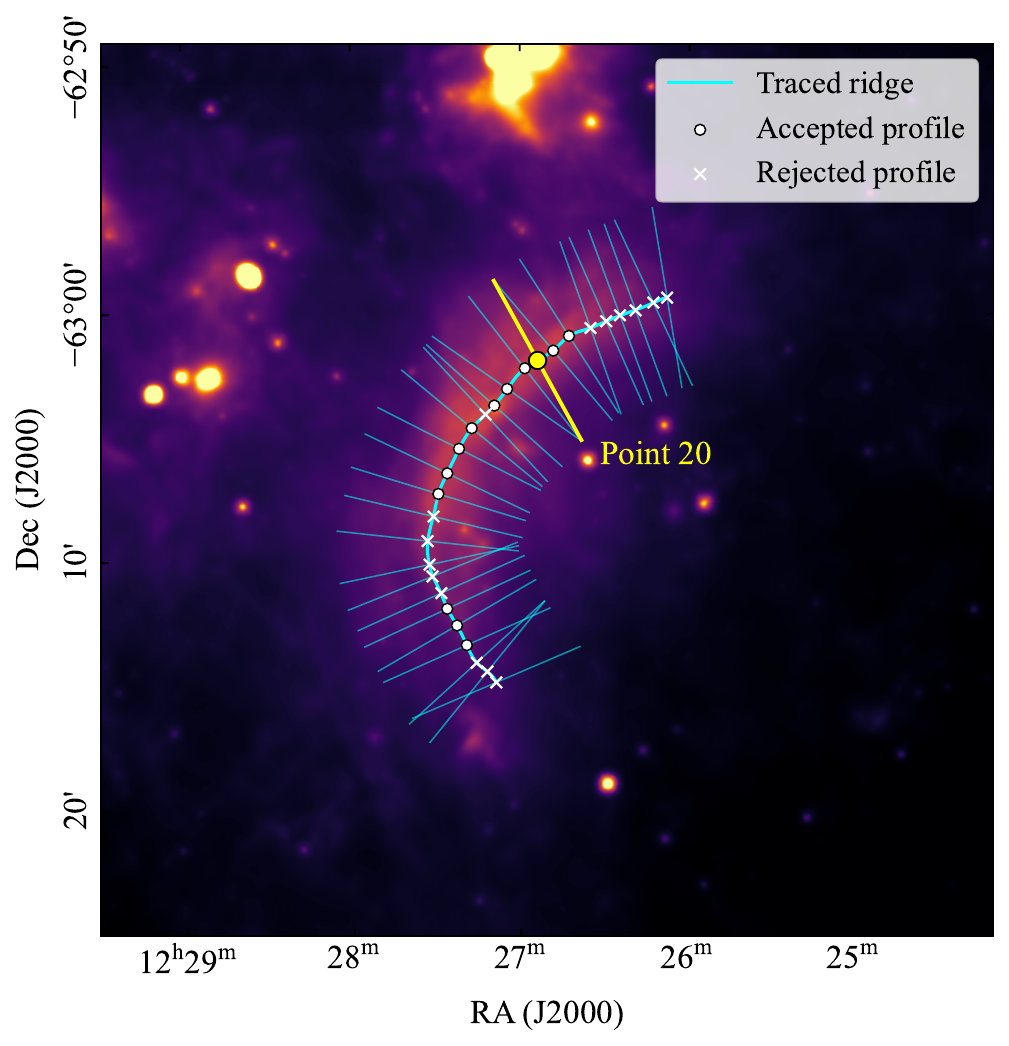}
\includegraphics[width=7.3cm]{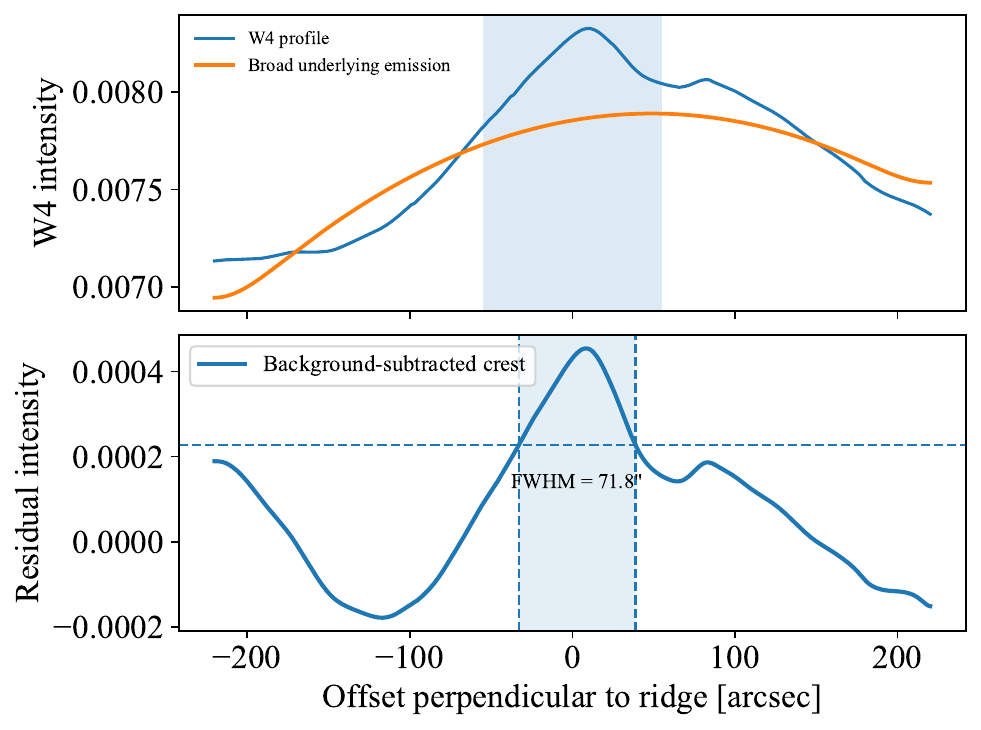}
\caption{\textit{Left panel:} \textit{WISE} W4 image of the infrared arc with the thickness measurement geometry. The cyan curve shows the traced bright ridge. Perpendicular cuts are used to measure the local ridge thickness. White points denote the ridge positions for which reliable width measurements were obtained, while white crosses indicate rejected cuts. The yellow segment highlights one representative accepted cut. \textit{Right panel:} Representative accepted perpendicular profile (Point~20). The upper plot shows the original W4 intensity profile (blue) and the estimated broad underlying emission (yellow). The lower plot shows the residual ridge profile after subtraction of this diffuse component. The dashed lines mark the half-maximum level and the FWHM boundaries, giving an observed width of 71.8~arcsec.}
\label{fig:thickness}
\end{figure}

To examine the environment of the system on larger angular scales, we also inspected \textit{Planck} maps at 30, 44, and 70~GHz \citep{Planck20}. We used the \textit{eROSITA} eRASS1 catalogue \citep{Merloni2024} fluxes of the two positional X-ray counterparts associated with the system and treated the HD~108248/HD~108249 and HD~108250 detections separately. The X-ray luminosities were calculated from the catalogue fluxes as
\[
L_{\rm X}=4\pi d^{2}F_{\rm X},
\]
where $F_{\rm X}$ is the eRASS1 flux in the 0.2--2.3~keV band and $d=106.4$~pc is the adopted distance to the system.

\subsection{The geometry of the W4  structure}
\label{geo}
To quantify the geometry of the W4  structure, we selected 28 points along the bright crest of the arc and fitted a smooth spline through them. At each ridge position, an intensity profile was extracted perpendicular to the local direction of the fitted curve. Each profile extended by $\pm$220~arcsec from the ridge and was sampled at 1~arcsec intervals.

Because the ridge is superposed on broader diffuse infrared emission, the measured profiles contain both the narrow bright crest and a more slowly varying underlying component. To isolate the crest, the central $\pm$55~arcsec region was excluded while estimating the broad underlying emission from the outer parts of each profile. This smooth component was then subtracted from the original profile, and the full width at half maximum (FWHM) of the remaining crest was measured directly from its two half-maximum crossings, without assuming a particular functional form for the profile. Figure~\ref{fig:thickness} illustrates the procedure for a representative accepted cut.

Only profiles with a clearly identified central crest and a reliable width measurement were retained. We required a crest signal-to-noise ratio greater than 3, a peak position within 30\,arcsec of the traced ridge, an observed FWHM between 12\,arcsec and 120\,arcsec, and a ratio of the larger to the smaller half-width at half maximum of less than 3. Profiles were also rejected when the profile did not cross the half-maximum level on both sides of the crest peak, preventing a reliable FWHM measurement. After subtraction of the broad underlying emission, the observed FWHM of the residual crest was measured. The intrinsic crest width was then estimated by removing the adopted 12\,arcsec W4 point-spread function (PSF) in quadrature \citep{wright10},
\[
{\rm FWHM}_{\rm int}
=
\sqrt{{\rm FWHM}_{\rm obs}^{2}
-
{\rm FWHM}_{\rm PSF}^{2}}.
\]
The median intrinsic FWHM of the accepted profiles was adopted as the characteristic thickness of the W4 ridge.

For each star, we identified the nearest point on the traced W4 ridge and calculated the corresponding minimum projected separation. The resulting star--ridge direction is expected geometrically to be approximately perpendicular to the local ridge tangent and was therefore not interpreted as independent evidence for a bow-shock geometry.

\begin{figure}
\centering
\includegraphics[width=11.0cm]{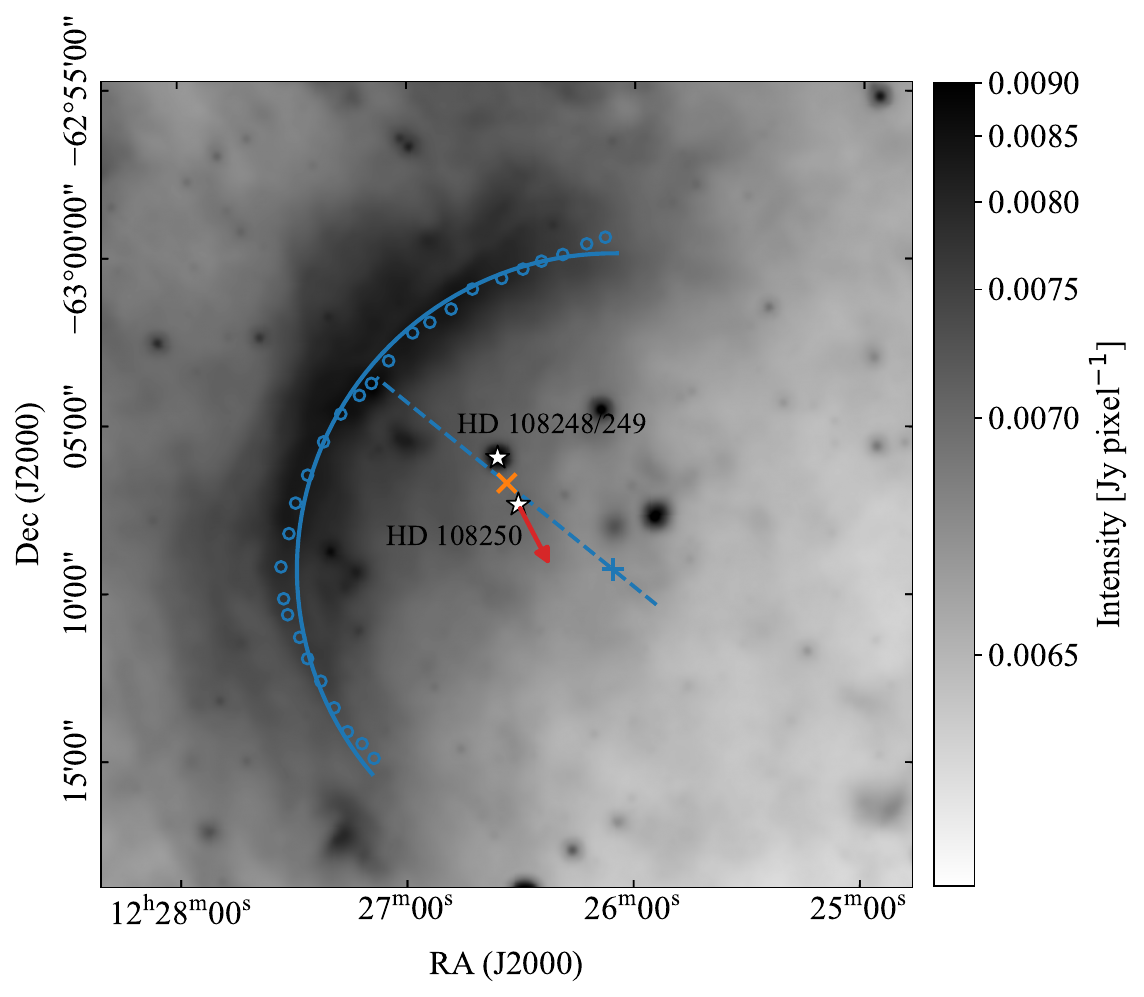}
\caption{Geometry of the infrared arc and the LCC-relative proper motion of HD~108250. The background shows the WISE W4 (22~$\mu$m) image. Blue open circles mark the 28 positions used to trace the bright W4 ridge, and the solid blue curve shows the best-fitting circular arc used to characterize its overall curvature. The blue cross marks the fitted centre of curvature, while the orange cross indicates the adopted stellar-system reference position. The dashed blue line shows the resulting geometrical symmetry axis. The white stars indicate the positions of the HD~108248/HD~108249 pair and HD~108250, and the red arrow shows the direction of the LCC-relative proper motion of HD~108250; its length is illustrative.}
\label{fig:arcgeo}
\end{figure}

The 28 positions selected along the bright W4 ridge were also used to characterize the overall curvature of the infrared structure (see Figure~\ref{fig:arcgeo}). A circle was fitted to these positions by minimizing the radial residuals of the ridge points from the fitted circle. The resulting circle was used as a simple geometrical characterization of the observed partial arc, and its centre defines the centre of curvature. For the geometrical comparison, we adopted a reference position for the stellar system at $\alpha_{\rm J2000}=12^{\rm h}26^{\rm m}33.619^{\rm s}$ and $\delta_{\rm J2000}=-63^\circ06'43.611''$. The geometrical symmetry axis was defined by the line connecting the fitted centre of curvature and this reference position. Position angles (PAs) were measured east of north. For the directional comparison, the axis PA was defined from the fitted curvature centre toward the stellar-system reference position; when continued through the reference position, this direction points toward the observed infrared arc. The orientation of this axis was subsequently compared with the LCC-relative proper-motion direction of HD~108250.

\subsection{Dust temperature and column-density analysis}
\label{maps}        
To characterize the physical conditions of the surrounding ISM, we used \textit{Herschel} continuum images at 160, 250, 350, and 500~$\mu$m. The 160~$\mu$m data were obtained with the PACS, while the 250, 350, and 500~$\mu$m observations were acquired with the SPIRE. Together, these bands sample the far-infrared spectral energy distribution of the cold dust and provide constraints on the dust temperature, $T_{\rm dust}$, and molecular hydrogen column density, $N(\mathrm{H}_2)$ \citep{Molinari2010}.

Assuming optically thin far-infrared dust emission, the observed specific intensity at each frequency was modelled with a modified blackbody of the form

\begin{equation}
I_{\nu} =
\kappa_{\nu}\,
\mu_{\mathrm{H}_2}m_{\rm H}\,
N(\mathrm{H}_2)\,
B_{\nu}(T_{\rm dust}),
\label{eq:SED}
\end{equation}
where $\kappa_{\nu}$ is the dust opacity per unit total gas-plus-dust mass, $\mu_{\mathrm{H}_2}=2.8$ is the mean molecular weight per hydrogen molecule \citep{Kauffmann2008}, $m_{\rm H}$ is the hydrogen-atom mass, and $B_{\nu}(T_{\rm dust})$ is the Planck function. The frequency dependence of the dust opacity was adopted as

\begin{equation}
\kappa_{\nu} =
\kappa_{0}
\left(\frac{\nu}{\nu_{0}}\right)^{\beta},
\label{eq:opacity}
\end{equation}
with $\kappa_{0}=0.1\,\mathrm{cm^{2}\,g^{-1}}$ at a reference wavelength of 300~$\mu$m and a fixed dust-emissivity index of $\beta=2$ \citep{Konyves2015}. This opacity law assumes a gas-to-dust mass ratio of 100.

Prior to the fitting, the 160, 250, and 350~$\mu$m images were convolved to the angular resolution of the SPIRE 500~$\mu$m data. All four maps were then reprojected onto a common grid with a pixel size of 14\,arcsec, corresponding to that of the 500~$\mu$m image. The modified-blackbody fit was performed independently for each pixel.

Following \citet{Pezzuto2021}, we constructed a grid of models with dust temperatures between 5 and 50~K in steps of 0.01~K and calculated the corresponding far-infrared intensities at each trial temperature. For a fixed temperature, the model intensity depends linearly on $N(\mathrm{H}_2)$; therefore, the column density was determined through least-squares scaling. For each pixel, the values of $T_{\rm dust}$ and $N(\mathrm{H}_2)$ corresponding to the best-fitting model were adopted. Calibration uncertainties of 20\% for the PACS 160~$\mu$m data and 10\% for each of the SPIRE bands were included in the fitting procedure \citep{Konyves2015}.

To compare the physical conditions toward the infrared arc with those of its surroundings, we used the region defined by the WISE W4 structure as the arc aperture. The local background was sampled using 17 regions distributed around the arc while avoiding the W4-defined structure and obvious compact sources. For each background region, the median value of the valid pixels was calculated separately in the $T_{\rm dust}$ and $N({\rm H_2})$ maps, and the median of these regional values was adopted as the representative local background. The corresponding median values within the W4-defined region were adopted for the arc.

\subsection{Astrometry and stellar motions}
\label{astrometry}
The coordinates of the three stars of $\alpha$~Crucis were taken from \textit{SIMBAD} astronomical database. HD~108248 and HD~108249 form a close pair with an angular separation of approximately 4~arcsec, whereas HD~108250 lies about 90~arcsec from the pair.

For HD~108250, we adopted the \textit{Gaia}~DR3 proper motion of ($\mu_{\alpha *},\mu_{\delta}$) = (--39.548$\pm$0.117,\,--13.760$\pm$0.132)\,mas\,yr$^{-1}$. HD~108248 and HD~108249 are too bright for reliable \textit{Gaia} astrometry, and we therefore used the proper motions from the Tycho Double Star Catalogue \citep{Fabricius2002}: 
(--35.4$\pm$0.6,\,--14.7$\pm$0.7)\,mas\,yr$^{-1}$ for HD~108248 and (--42.5$\pm$1.0,\,--7.7$\pm$1.0)\,mas\,yr$^{-1}$ for HD~108249.

The stellar motions were compared with the mean proper motion of the LCC subgroup, $(\mu_{\alpha *},\mu_{\delta})_{\rm LCC}$= (--37.91$\pm$0.89,\,--10.62$\pm$1.41)\,mas\,yr$^{-1}$, reported by \citet{Torosyan24}. First, the proper motion of each star was compared with the mean motion of LCC by defining
\[
\Delta\mu_{\alpha *}
=
\mu_{\alpha *,{\rm star}}-\mu_{\alpha *,{\rm LCC}},
\qquad
\Delta\mu_{\delta}
=
\mu_{\delta,{\rm star}}-\mu_{\delta,{\rm LCC}}.
\]
The magnitude of the LCC-relative proper-motion vector was then calculated as
\[
\Delta\mu =
\sqrt{
(\Delta\mu_{\alpha *})^2+
(\Delta\mu_{\delta})^2
}.
\]
The corresponding tangential peculiar velocity was obtained from
\[
v_{\rm t,pec}
=
4.74047\,
\frac{\Delta\mu}{1000}\,
d,
\]
where $\Delta\mu$ is in mas\,yr$^{-1}$, $d$ is the distance in pc, and $v_{\rm t,pec}$ is in km\,s$^{-1}$.  

The PA of the LCC-relative proper-motion vector was calculated from $\Delta\mu_{\alpha *}$ and $\Delta\mu_{\delta}$. For HD~108250, this direction was compared with the geometrical symmetry axis of the infrared arc defined in Sect.~\ref{geo} and shown in Figure~\ref{fig:arcgeo}. Because a symmetry axis has no intrinsic directional sense, the angular offset from the axis was evaluated modulo $180^\circ$; the directed orientation relative to the arc-facing side of the axis was considered separately.

Since HD~108248 and HD~108249 form a very close pair, we also considered their mean proper motion as a representative motion of the pair. Because the representative LCC-relative motion of the HD~108248/HD~108249 pair is consistent with zero within its uncertainties, its residual position angle is not meaningfully constrained and no directional comparison with the infrared arc was performed. The uncertainty in the LCC-relative motion of HD~108250 was propagated using Monte Carlo simulations. We generated $2\times10^{5}$ realizations, independently sampling the two proper-motion components of HD~108250 and the mean LCC motion from Gaussian distributions defined by their measured values and uncertainties. For each realization, the sampled LCC motion was subtracted from the stellar proper motion, and the magnitude and position angle of the resulting peculiar proper-motion vector were recalculated. The median and 16th--84th percentile ranges of the resulting distributions were adopted as the representative values and uncertainties. Correlations between the proper-motion components were not included in the sampling. The significance of the peculiar motion was additionally evaluated with a two-dimensional $\chi^2$ test for zero LCC-relative motion, assuming independent uncertainties in the two proper-motion components. For each realization, we also calculated the angular offset between the resulting peculiar-motion direction and the geometrical symmetry axis of the arc, treating the axis modulo $180^\circ$.

\section{Results}
\label{sec:results}

\subsection{Infrared morphology and geometry of the arc}
\label{sec:arc_geometry}

\begin{figure}
\centering
    \includegraphics[width=11cm]{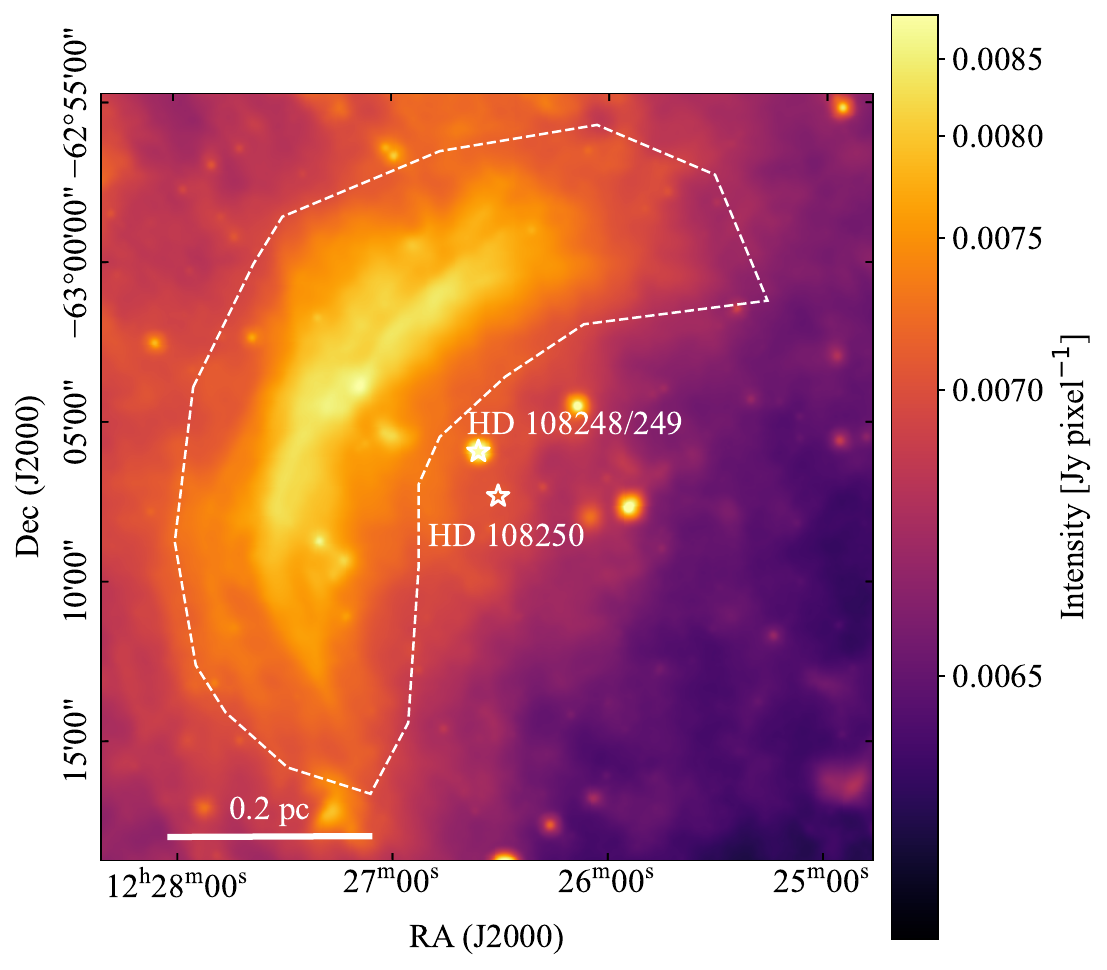}
    \caption{\textit{WISE} W4 ($22\,\mu$m) image of the region surrounding the $\alpha$~Crucis system. The white stars indicate the positions of the close HD~108248/HD~108249 pair and HD~108250. The dashed outline marks the prominent arc-like infrared structure investigated in this work. The scale bar indicates a projected length of 0.2~pc at the adopted distance of 106.4~pc.}
    \label{fig:wise}
\end{figure}

The \textit{WISE} W4 image reveals a prominent curved infrared structure adjacent to the $\alpha$~Crucis system. Figure~\ref{fig:wise} shows the arc-like structure to the east of the system. The emission forms a well-defined bright ridge over a substantial fraction of the structure, while the surrounding field contains more diffuse and irregular mid-infrared emission.

\begin{figure}
\includegraphics[width=6.5cm]{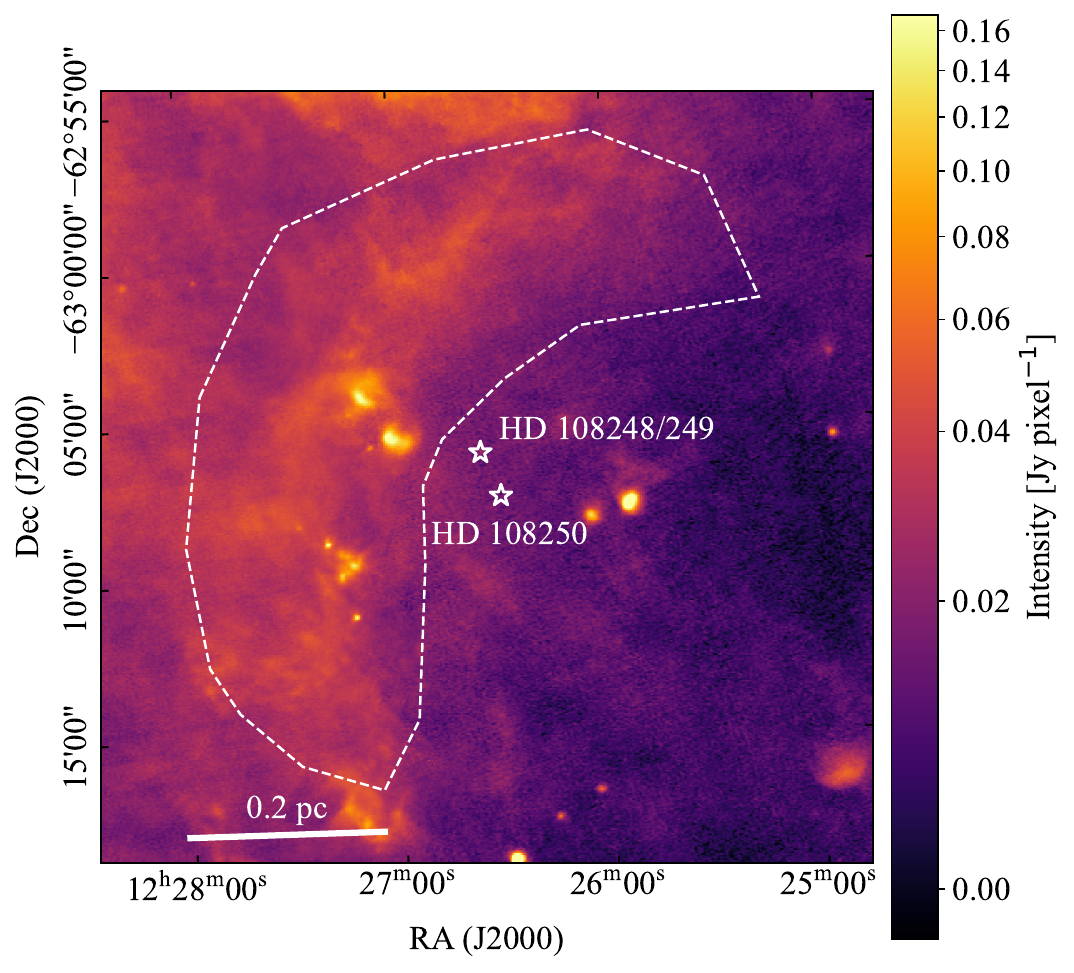}
\includegraphics[width=6.5cm]{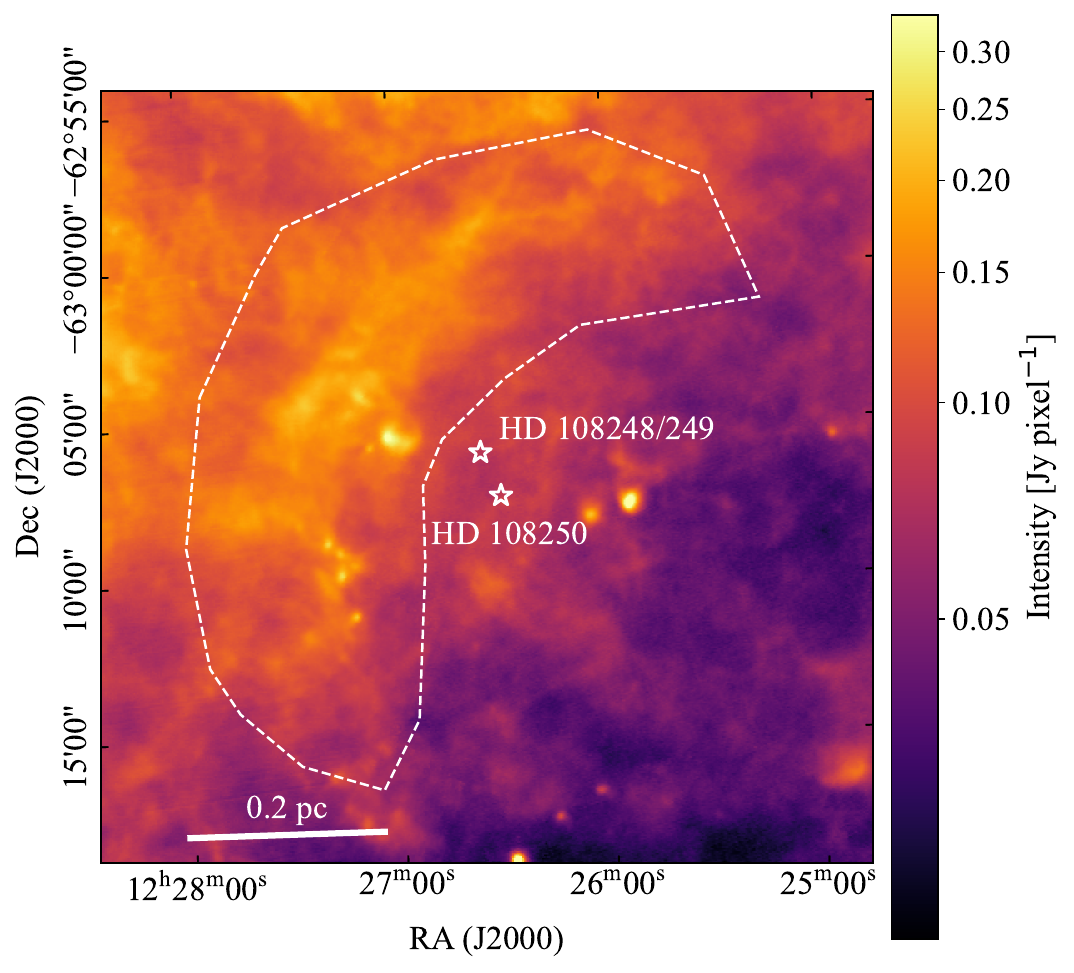}
\caption{\textit{Herschel}~PACS images of the region surrounding the $\alpha$~Crucis system at 70~$\mu$m (\textit{left panel}) and 160~$\mu$m (\textit{right panel}). The white stars indicate the positions of the close HD~108248/HD~108249 pair and HD~108250. The dashed outline marks the region occupied by the prominent infrared structure identified in the \textit{WISE} W4 image (Fig.~\ref{fig:wise}). The scale bar corresponds to 0.2~pc at the adopted distance of 106.4~pc.}
\label{fig:herschel}
\end{figure}

The appearance of the region changes considerably toward longer wavelengths. Figure~\ref{fig:herschel} shows the corresponding \textit{Herschel}~PACS images at 70 and 160~$\mu$m. At 70~$\mu$m, diffuse and structured far-infrared emission is present within the W4-defined region, together with several compact bright sources. The narrow, continuous morphology of the 22~$\mu$m ridge, however, is much less clearly defined at 70~$\mu$m. At 160~$\mu$m, the contrast of the arc decreases further. Rather than a sharply defined curved ridge, the emission is dominated by a broader and spatially structured far-infrared component extending over the same general region. Thus, the narrow W4 morphology is not reproduced as an equally well-defined structure at longer wavelengths. The wavelength-dependent morphology suggests that the narrow 22~$\mu$m ridge traces a warmer dust component associated with a more extended distribution of cooler ISM.

To quantify the geometry of the W4 structure, we traced the bright ridge and measured its projected extent and thickness. The traced ridge extends over approximately 21.1~arcmin. At the adopted distance of $106.4\pm1.5$~pc, this corresponds to a projected length of approximately 0.65~pc.

Of the 28 perpendicular profiles sampled along the ridge, 13 satisfied the adopted quality criteria and were retained for the width measurement. The accepted profiles give a median observed FWHM of 71.8~arcsec. After correction for the W4 PSF, the median intrinsic FWHM is 70.8~arcsec, corresponding to approximately 0.037~pc. The intrinsic widths vary along the traced ridge, with a dispersion of approximately 24~arcsec ($\sim0.012$~pc), indicating that the thickness is not completely uniform along the arc.

The projected minimum separations from the stellar components to the traced ridge are 259.1\,arcsec for HD~108248, 257.0\,arcsec for HD~108249, and 337.3\,arcsec for HD~108250. Thus, the close HD~108248/HD~108249 pair lies systematically closer to the infrared ridge than HD~108250. Taking the mean separation of the close pair gives 258.1\,arcsec, or approximately $0.133$~pc, whereas the corresponding projected separation for HD~108250 is approximately $0.174$~pc. To examine whether the bright W4 ridge is more nearly concentric with one of the stellar components, we calculated the angular distance $r_i$ from each star to each of the 28 points used to trace the ridge. For each candidate stellar centre, we computed the mean radius $\langle r\rangle$ and its standard deviation $\sigma_r$, and used $\sigma_r/\langle r\rangle$ as a simple measure of the fractional variation in radius. A smaller value therefore corresponds to a more nearly constant-radius ridge about that stellar position.

The resulting fractional radial dispersions are $\sigma_r/\langle r\rangle=0.266$ for HD~108248, $0.265$ for HD~108249, and $0.138$ for HD~108250. Thus, the traced W4 ridge is substantially closer to a constant-radius curve centred on HD~108250 than on either member of the close pair. As a robust check less sensitive to individual outlying points, we also calculated the median absolute deviation, ${\rm MAD}=\mathrm{median}(|r_i-\mathrm{median}(r)|)$, normalized by the median radius. The resulting ${\rm MAD}/\mathrm{median}(r)$ values are 0.237, 0.237, and 0.119 for HD~108248, HD~108249, and HD~108250, respectively.  This geometrical comparison should nevertheless be treated cautiously because only a partial arc is observed. As a complementary characterization of the overall curvature, we fitted a circle to the same 28 ridge positions. The fit yields a centre of curvature at approximately $\alpha_{\rm J2000}=12^{\rm h}26^{\rm m}05.61^{\rm s}$, $\delta_{\rm J2000}=-63^\circ09'17.8''$, with a radius of $\sim565''$. The corresponding geometrical symmetry axis has a position angle of $PA_{\rm arc}=51.0^\circ$.

\subsection{Stellar kinematics relative to the infrared arc}
\label{sec:kinematics}

To investigate whether the infrared morphology is related to the motion of the stellar system, we compared the proper motions of the three stellar components with the mean motion of the LCC subgroup of the Scorpius--Centaurus association, $(\mu_{\alpha *},\mu_{\delta})_{\rm LCC}$ =(--37.91$\pm$0.89,\,--10.62$\pm$1.41)\,mas\,yr$^{-1}$ \citep{Torosyan24}. For HD~108250, we adopted the \textit{Gaia}\,DR3 proper motion, whereas for the bright HD~108248 and HD~108249 components we used measurements from the Tycho Double Star Catalogue. The astrometric and kinematic properties of the individual stars are summarized in Table~\ref{tab:stars}. The table also lists their projected peculiar tangential velocities and minimum projected distances from the traced W4 ridge.

\begin{table*}
\centering
\caption{Astrometric, kinematic, and geometric properties of the stellar components.}
\resizebox{1\textwidth}{!}{
\label{tab:stars}
\begin{tabular}{lcccccc}
\hline
Object & Sp. type & $\mu_{\alpha *}$ & $\mu_{\delta}$ & $|\Delta\mu|$ & $v_{\rm t,pec}$ & $d_{\rm ridge}$ \\
& & (mas yr$^{-1}$) & (mas yr$^{-1}$) & (mas yr$^{-1}$) & (km s$^{-1}$) & (pc) \\
\hline
HD~108248 & B0.5 IV & $-35.4\pm0.6$ & $-14.7\pm0.7$ & 4.79 & $2.42\pm0.73$ & 0.134 \\
HD~108249 & B1 V & $-42.5\pm1.0$ & $-7.7\pm1.0$ & 5.44 & $2.74\pm0.74$ & 0.133 \\
HD~108250 & B4 V & $-39.548\pm0.117$ & $-13.760\pm0.132$ & 3.54 & $1.79\pm0.67$ & 0.174 \\
\hline
\end{tabular}
}
\begin{flushleft}
\footnotesize
\textbf{Notes.} The peculiar proper motions are measured relative to the mean motion of the LCC subgroup. The proper motions of HD~108248 and HD~108249 are from the Tycho Double Star Catalogue \citep{Fabricius2002}, while that of HD~108250 is from \textit{Gaia}~DR3 \citep{Gaia22}. The individual residual motions of HD~108248 and HD~108249 are listed for completeness; because they form a close pair, their representative combined motion is used in the interpretation. The tangential peculiar velocities assume a distance of 106.4~pc. $d_{\rm ridge}$ is the minimum projected separation between each stellar position and the traced WISE W4 ridge, converted to physical distance using the adopted distance of 106.4 pc.
\end{flushleft}
\end{table*}

The individual Tycho proper motions of HD~108248 and HD~108249 differ noticeably. Because the two stars form a close pair separated by only $\sim$4~arcsec, we do not interpret their individual LCC-relative motions separately. Instead, we adopted their arithmetic mean as a representative proper motion of the close pair. For each proper-motion component, the uncertainty was estimated by combining in quadrature the propagated uncertainty of the mean and the scatter of the two measurements about their mean. This gives $(\mu_{\alpha*},\mu_{\delta})_{\rm pair} =(-38.95\pm3.60,-11.20\pm3.55)$~mas~yr$^{-1}$. Relative to the mean LCC motion, this gives $(\Delta\mu_{\alpha*},\Delta\mu_\delta)_{\rm pair} =(-1.04\pm3.71,-0.58\pm3.82)$~mas\,yr$^{-1}$, with a vector magnitude of approximately $1.2$~mas\,yr$^{-1}$. The pair motion is therefore consistent with that of LCC within the uncertainties, and the direction of its residual motion is not meaningfully constrained.

HD~108250 shows a different behaviour. After subtraction of the mean LCC motion, we obtain $(\Delta\mu_{\alpha *},\Delta\mu_{\delta})$=(--1.638$\pm$0.898,\,--3.140$\pm$1.416)\,mas\,yr$^{-1}$, corresponding to a nominal residual amplitude of $|\Delta\mu|$ = 3.54~mas\,yr$^{-1}$. Using the adopted distance of 106.4~pc and the relation given in Sect.~\ref{astrometry}, this residual proper-motion amplitude corresponds to a projected peculiar tangential velocity of $v_{\rm t,pec}=1.79\pm0.67$~km~s$^{-1}$.

The orientation of the peculiar-motion vector was then compared with the geometry of the infrared structure. Figure~\ref{fig:arcgeo} shows the traced W4 ridge together with the stellar positions and the LCC-relative proper-motion vector of HD~108250. The LCC-relative proper-motion vector of HD~108250 has a position angle of $207.6^\circ$, whereas the geometrical symmetry axis of the infrared arc has $PA_{\rm arc}=51.0^\circ$. The directed angular separation is therefore $156.6^\circ$. Treating the geometrical axis modulo $180^\circ$ gives an axis misalignment of $23.4^\circ$. Thus, although the proper motion lies relatively close to the geometrical symmetry axis, it points in the opposite sense from the infrared arc. This is not the orientation expected for a simple motion-driven bow shock whose upstream apex lies in the direction of the stellar motion.

For HD~108250, Monte Carlo propagation of the astrometric uncertainties yields a median residual amplitude of $3.68^{+1.29}_{-1.23}$~mas~yr$^{-1}$. A two-dimensional test gives $\chi^2=8.25$ for two degrees of freedom, corresponding to $p=0.016$, indicating a statistically significant but low-amplitude departure from the mean LCC motion.

Monte Carlo propagation gives a median symmetry-axis offset of 24.6$^\circ$, with a 16th--84th percentile range of 9.5$^\circ$--39.2$^\circ$. Approximately 64\% of the realizations lie within 30$^\circ$ of the symmetry axis.

Thus, neither the HD~108248/HD~108249 pair nor HD~108250 exhibits a large peculiar transverse velocity relative to LCC. The close pair is consistent with the motion of the association, whereas HD~108250 shows a low-amplitude but statistically significant peculiar motion that lies relatively close to the geometrical symmetry axis but points in the opposite sense from the infrared arc.

\subsection{Dust temperature and column density}
\label{sec:dust}

\begin{figure}
\includegraphics[width=7cm]{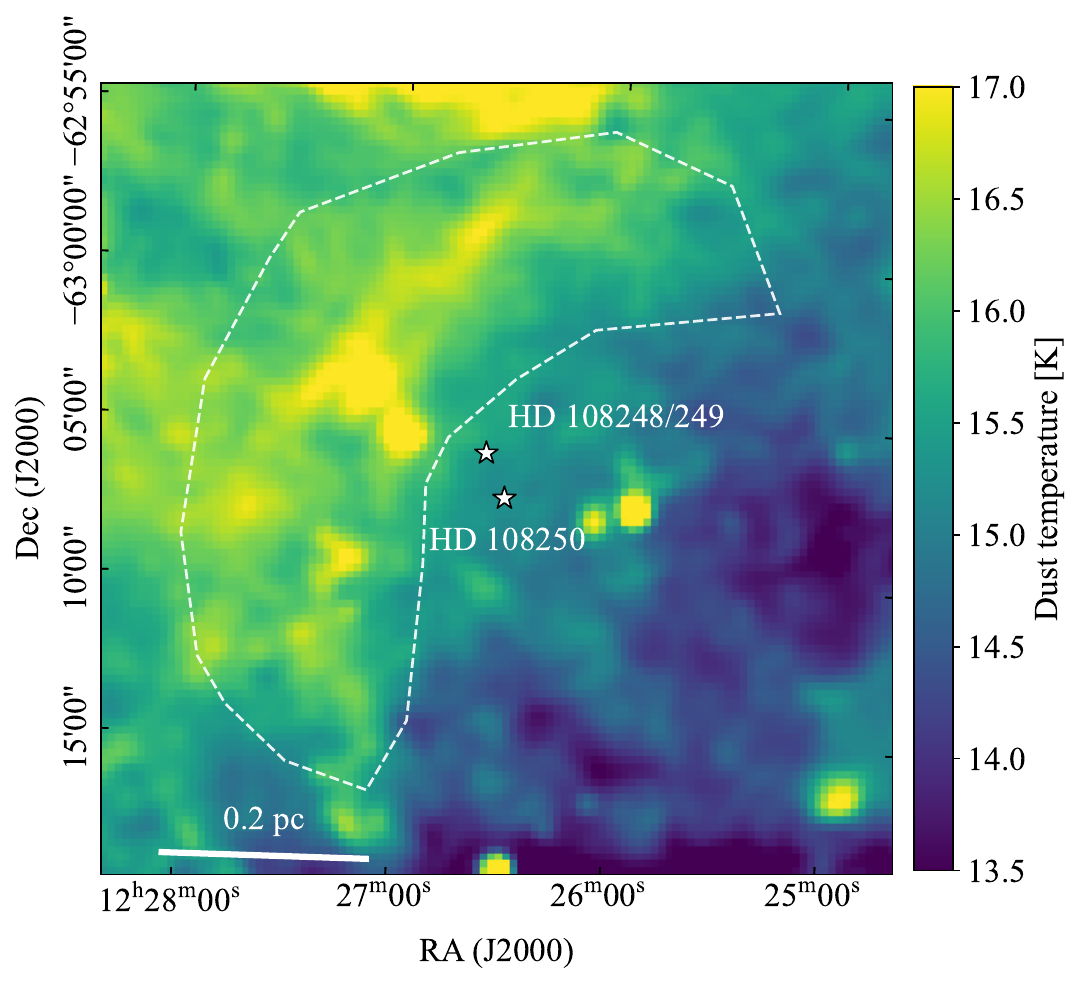}
\includegraphics[width=7cm]{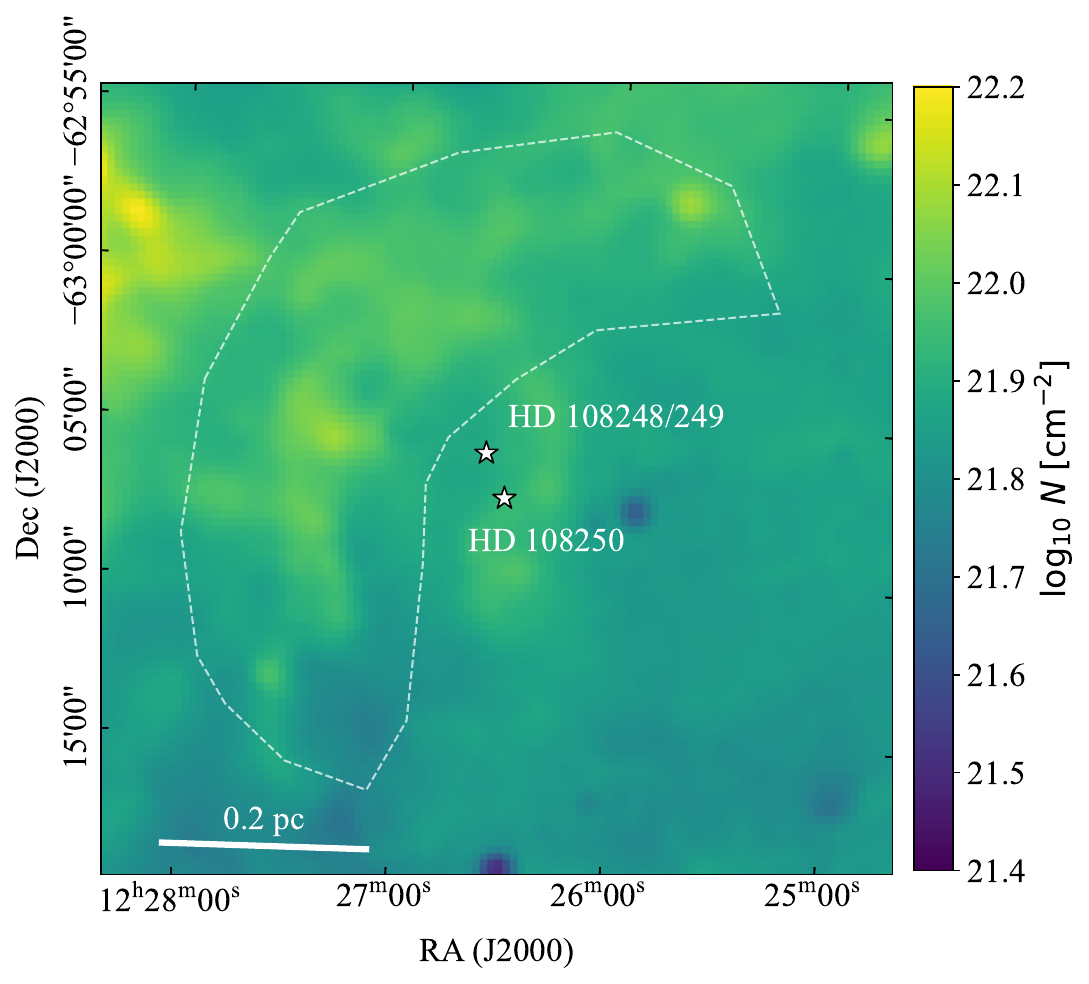}
\caption{Dust properties of the region surrounding the $\alpha$~Crucis system derived from the \textit{Herschel} far-infrared data. \textit{Left panel}: dust-temperature map. \textit{Right panel}: molecular-hydrogen column-density map, shown as $\log_{10}[N({\rm H_2})/{\rm cm^{-2}}]$. The white stars mark the positions of the close HD~108248/HD~108249 pair and HD~108250. The dashed outline indicates the region occupied by the infrared structure identified in the \textit{WISE} W4 image. The scale bar indicates a projected length of 0.2~pc at the adopted distance of 106.4~pc.}
\label{fig:maps}
\end{figure}

Figure~\ref{fig:maps} shows the dust-temperature and molecular-hydrogen column-density distributions across the region. Both quantities display substantial spatial structure, although the narrow morphology of the W4 ridge is not reproduced as a similarly well-defined feature in either map.

The region associated with the W4 arc has a  characteristic dust temperature of $T_{\rm d}\simeq16.13$~K, compared with $T_{\rm d}\simeq15.33$~K in the nearby background, corresponding to an increase of approximately 0.8~K. The molecular-hydrogen column density shows a similar behaviour, with $N({\rm H_2})_{\rm arc} \simeq$8.44$\times$10$^{21}$~cm$^{-2}$, compared with $N({\rm H_2})_{\rm bg}\simeq$7.72$\times$10$^{21}$~cm$^{-2}$. The corresponding column-density difference is approximately $7.2\times10^{20}$~cm$^{-2}$.

Thus, the material projected toward the W4-defined region is slightly warmer and has a somewhat higher column density than the neighbouring background. However, these differences form part of a broader and spatially structured far-infrared environment rather than a sharply confined feature coincident with the narrow 22~$\mu$m ridge. 

The main geometric and far-infrared properties of the arc are summarized in Table~\ref{tab:arc}. The table combines the characteristic dimensions derived from the W4 ridge with the dust temperature and molecular-hydrogen column density measured toward the arc and the nearby background.

Overall, the arc is a narrow mid-infrared structure embedded within a broader far-infrared environment. The median dust temperature and column density within the W4-defined region are only slightly higher than those of the surrounding material. We therefore treat these differences as small arc--background contrasts and do not assign them a formal statistical significance.

\begin{table}
\centering
\caption{Summary of the measured geometric and far-infrared properties of
the infrared arc and its local environment.}
\label{tab:arc}
\begin{tabular}{ll}
\hline
Quantity & Value \\
\hline
Adopted distance & $106.4\pm1.5$ pc \\
Traced ridge length & 21.1\,arcmin ($0.654$ pc) \\
Median observed FWHM & 71.76\,arcsec \\
Median intrinsic FWHM & 70.75\,arcsec ($0.0365$ pc) \\
Intrinsic FWHM dispersion & 24.03\,arcsec ($0.0124$ pc) \\
HD~108248/249--ridge separation & 258.1\,arcsec ($0.133$ pc) \\
HD~108250--ridge separation & 337.3\,arcsec ($0.174$ pc) \\
$T_{\rm dust,arc}$ & 16.13 K \\
$T_{\rm dust,bg}$ & 15.33 K \\
$\Delta T_{\rm dust}$ & 0.80 K \\ 
$N_{\rm H_2,arc}$ & $8.44\times10^{21}$ cm$^{-2}$ \\
$N_{\rm H_2,bg}$ & $7.72\times10^{21}$ cm$^{-2}$ \\
$\Delta N_{\rm H_2}$ & $7.2\times10^{20}$ cm$^{-2}$ \\
\hline
\end{tabular}
\end{table}

\begin{figure}
\centering
\includegraphics[width=11cm]{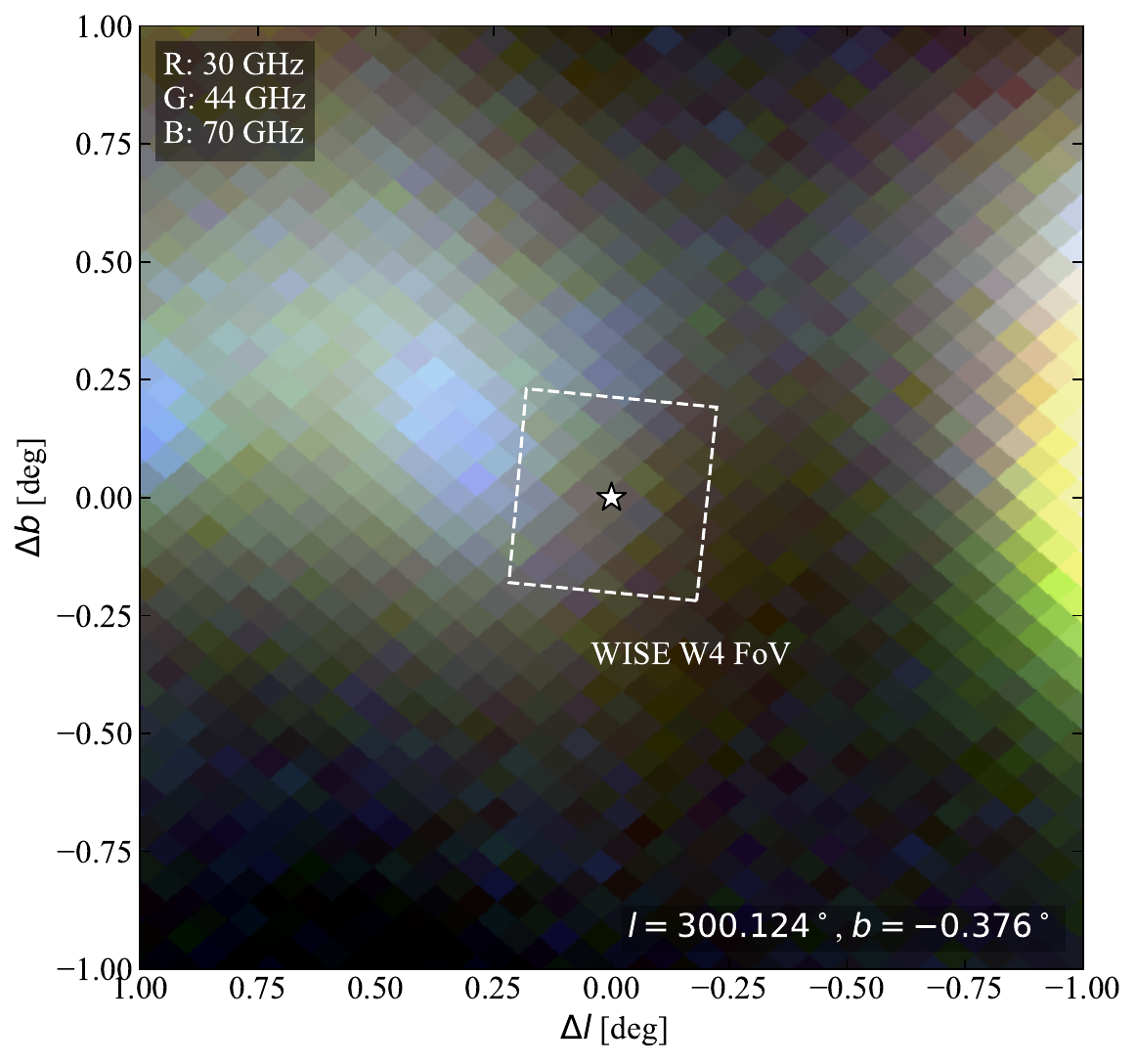}
\caption{Large-scale \textit{Planck} microwave environment of the $\alpha$~Crucis system. The RGB composite combines the \textit{Planck} 30~GHz (red), 44~GHz (green), and 70~GHz (blue) maps. The white star marks the position of the stellar system, and the dashed rectangle indicates the field covered by the \textit{WISE} W4 image shown in Fig.~\ref{fig:wise}. Coordinates are shown as offsets in Galactic longitude and latitude from $(l,b)=(300.124^\circ,-0.376^\circ)$. The \textit{Planck} image is used to illustrate the large-scale microwave environment and does not resolve the infrared arc itself.}
\label{fig:planck}
\end{figure}

\subsection{Large-scale microwave environment}
\label{sec:planck}
The environment of the system on degree scales is shown in Figure~\ref{fig:planck}, using an RGB combination of the \textit{Planck} 30, 44, and 70~GHz maps. The W4 field occupies only a small region within a substantially larger and spatially structured microwave background. Pronounced large-scale intensity gradients and spatial variations are visible across the field, showing that the stellar system is projected against a non-uniform microwave environment.

Because of the much lower angular resolution of the \textit{Planck} data, the infrared ridge itself cannot be resolved at these frequencies. Moreover, emission in the 30--70~GHz range may contain contributions from several mechanisms, including synchrotron, free--free, and anomalous microwave emission \citep{Planck2016XXV}. We therefore use these maps primarily to characterize the large-scale environmental structure rather than to identify the physical origin of the W4 arc.

\subsection{X-ray emission}
\label{sec:xray}
Two distinct \textit{eROSITA} eRASS1 sources are positionally associated with the stellar system. The source 1eRASS~J122636.1$-$630558 corresponds to the close HD~108248/HD~108249 pair, whereas 1eRASS~J122630.8$-$630718 is associated with HD~108250. Their principal X-ray properties are summarized in Table~\ref{tab:xray}.

The X-ray emission associated with the HD~108248/HD~108249 pair is considerably stronger than that of HD~108250. In the 0.2--2.3~keV band, the pair has a flux of $\sim1.3\times10^{-11}$~erg~s$^{-1}$~cm$^{-2}$, compared with $\sim2.2\times10^{-13}$~erg~s$^{-1}$~cm$^{-2}$ for HD~108250. At the adopted distance, these values correspond to luminosities of approximately $1.8\times10^{31}$ and $3.0\times10^{29}$~erg~s$^{-1}$, respectively. Thus, the HD~108248/HD~108249 source is about 60 times more luminous in X-rays than HD~108250.

The \textit{eROSITA} detection associated with the close pair cannot be assigned uniquely to HD~108248 or HD~108249 at the available angular resolution, and we therefore treat it as the combined X-ray counterpart of the pair.

\begin{table}
\centering
\caption{\textit{eROSITA} eRASS1 X-ray sources associated with the $\alpha$~Crucis system.}
\label{tab:xray}
\begin{tabular}{lcccc}
\hline
Counterpart & DET\_LIKE & Count rate & $F_{\rm X}$ & $L_{\rm X}$ \\
& & (counts s$^{-1}$) & (erg s$^{-1}$ cm$^{-2}$) & (erg s$^{-1}$) \\
\hline
HD~108248/249 & 22381.6 & 0.2446 & $1.30\times10^{-11}$ & $1.8\times10^{31}$ \\
HD~108250 & 65.1 & 0.0369 & $2.22\times10^{-13}$ & $3.0\times10^{29}$ \\
\hline
\end{tabular}
\begin{flushleft}
\footnotesize
\textbf{Notes.} The X-ray fluxes refer to the 0.2--2.3~keV band, and the luminosities were calculated assuming a distance of 106.4~pc. DET\_LIKE is the source-detection likelihood reported in the eRASS1 catalogue; larger values correspond to more significant detections \citep{Merloni2024}.
\end{flushleft}
\end{table}

\section{Discussion}
\label{sec:discussion}

\subsection{Nature of the infrared arc}
\label{sec:discussion_arc}

The morphology of the W4 emission initially suggests a possible wind bow-shock-like structure associated with the $\alpha$~Crucis system. Mid-infrared arcs around OB stars are commonly identified as candidate stellar bow shocks, with large surveys revealing several hundred such structures at 22--24~$\mu$m \citep{kobulnicky16,Kobulnicky2017}. The traced ridge length of $\sim$0.65~pc is relatively small compared with many of the bow shocks reported by \citet{Carretero-Castrillo2025}, although it remains within the range found in their sample. We note that our measured width of $\sim$0.037~pc refers to the FWHM thickness of the bright W4 ridge and is therefore not directly comparable to the overall projected bow-shock width reported by \citet{Carretero-Castrillo2025}. However, the combination of the infrared morphology and the stellar kinematics in the present system does not provide a straightforward classical wind bow-shock interpretation.

The HD~108248/HD~108249 pair lies closest to the bright W4 ridge, at a projected separation of approximately 0.13~pc. Both stars are early B-type objects and are therefore expected to provide stronger stellar-wind input than the later-type HD~108250. Observational estimates for B0--B3 stars associated with infrared bow shocks indicate mass-loss rates of order $10^{-9}$--$10^{-8}\,M_{\odot}\,{\rm yr^{-1}}$ \citep{Kobulnicky2019}, demonstrating that even early-B stars can produce detectable infrared wind--ISM structures despite their relatively weak winds.

Nevertheless, when the motion of the close pair is considered as a whole, its proper motion is consistent with the mean motion of LCC. There is therefore no evidence for a substantial transverse peculiar velocity that would naturally produce a conventional motion-driven wind bow shock. HD~108250 presents a different situation. Its projected separation from the ridge is larger, approximately 0.17~pc, but its LCC-relative proper motion is measurably different from zero. Geometrically, the proper-motion vector lies relatively close to the symmetry axis of the infrared arc, with an axis misalignment of about $23^\circ$. However, the direction of motion is opposite to the arc-facing direction of the axis. Thus, the geometry does not correspond to that expected for a simple motion-driven bow shock in which the infrared apex lies upstream of the stellar motion.

The corresponding projected peculiar velocity of HD~108250, however, is only $\sim1.8$~km~s$^{-1}$, far below the velocities traditionally associated with classical runaway stars \citep[e.g., ][]{Carretero-Castrillo2023}. A low stellar peculiar velocity does not necessarily exclude the formation of an infrared wind--ISM interaction structure. In a recent study of the Vel~OB1 association, \citet{azatyan26} identified several probable bow shocks around stars with peculiar tangential velocities below the $15$~km~s$^{-1}$ threshold adopted for the kinematic selection of runaways. Some of these objects have velocities below the $\sim10$--15~km~s$^{-1}$ characteristic ISM sound speed adopted in that study, implying that the stellar motion alone may be insufficient to provide the required supersonic relative velocity.

\citet{azatyan26} found that the lower-velocity systems span a much wider range of misalignment angles between the stellar motion and the infrared-arc axis, whereas the higher-velocity systems tend to be better aligned, although this trend did not reach formal statistical significance because of the limited sample size. Such behaviour is expected when the relevant quantity is the relative velocity between the star and the ambient gas rather than the stellar velocity alone. At low stellar velocities, bulk flows and density structure in the ISM can become dynamically important and substantially modify the orientation of the interaction region \citep{Mackey2015,mackey16,Bodensteiner18}.

The very small transverse peculiar velocity of HD~108250 should therefore not, by itself, be taken as evidence against a wind--ISM interaction. However, the directional mismatch described above argues against the simplest motion-driven bow-shock geometry and suggests that local gas motions, density gradients, or pre-existing interstellar structure may contribute to the observed morphology. This interpretation is consistent with the recent statistical study of \citet{vandenEijnden2026}, who found that substantial misalignments between stellar peculiar motions and infrared bow-shock orientations are common, and that bulk ISM motions can be comparable to or dominate the stellar motion in many systems. They showed that characteristic ISM velocities of order 10--15~km~s$^{-1}$ can account for much of the observed orientation distribution. In the $\alpha$~Crucis system, where the transverse peculiar velocity of HD~108250 is only $\sim1.8$~km~s$^{-1}$, even modest local gas motions could therefore dominate the relative star--ISM velocity. In this respect, the structure could be similar to the so-called ``in situ'' bow shocks, in which the relative star--ISM motion is produced primarily by bulk motion of the surrounding gas rather than by a high stellar peculiar velocity \citep{Povich2008,kobulnicky16}. However, in the present case no specific large-scale gas flow has been identified around $\alpha$~Crucis to confirm this possibility. Radiation-hydrodynamic calculations likewise show that strongly asymmetric wind bubbles can develop around slowly moving massive stars. In the models of \citet{Mackey2015}, stellar velocities as low as $4$~km~s$^{-1}$ were sufficient to produce markedly asymmetric wind bubbles. Synthetic infrared observations of these models further showed that the edge of such a bubble can appear as a prominent 24~$\mu$m arc even in cases where no genuine stellar bow shock is present \citep{mackey16}. Unless a substantial unmeasured radial component is present, however, HD~108250 remains far from the classical high-velocity runaway regime.

The wavelength dependence of the infrared morphology provides a further constraint. The narrow W4 ridge becomes substantially less distinct at 70~$\mu$m and is no longer clearly identifiable as an individual structure at 160~$\mu$m. A similar trend is common among mid-infrared-selected stellar bow-shock candidates. In the sample of \citet{Kobulnicky2017}, 422 of 709 objects selected at 22--24~$\mu$m were detected at 70~$\mu$m, whereas only 39 were detected at 160~$\mu$m. The much stronger contrast of the present arc at 22~$\mu$m is therefore not unusual for an infrared structure heated by an early-type star.

There is, however, an important difference in the inferred dust properties. \citet{Kobulnicky2017} derived characteristic colour temperatures of approximately 80--400~K from the 24/70~$\mu$m ratios and 40--200~K from the 70/160~$\mu$m ratios of candidate bow shocks. In contrast, our 160--500~$\mu$m modified-blackbody analysis yields $T_{\rm dust}\simeq16$~K toward the W4-defined region, only $\sim0.8$~K higher than the representative nearby background. These temperatures are not directly comparable, because our fit is dominated by the colder large-grain component sampled at longer wavelengths, whereas the shorter-wavelength colour temperatures preferentially trace warmer dust close to the interaction region. The comparison therefore indicates only a small difference in the cold-dust component between the W4-defined region and its surroundings; it does not exclude the presence of a warmer dust component responsible for the prominent 22~$\mu$m emission.

The molecular-hydrogen column density likewise increases by only $\sim7\times10^{20}$~cm$^{-2}$ relative to the local background. The far-infrared maps therefore point to a comparatively modest perturbation of the cold ISM and do not reveal the pronounced column-density enhancement that might be expected from a strongly compressed shell. This does not exclude unresolved compression on smaller spatial scales, while numerical models show that instabilities, density gradients, and projection effects can produce considerably more complex structures than the simple analytic bow-shock geometry \citep{Meyer2014,Acreman2016}.

On considerably larger angular scales, the \textit{Planck} 30, 44, and 70~GHz maps reveal a spatially structured microwave environment around the position of the stellar system. The W4 field occupies only a small part of this extended structure, indicating that the infrared arc is projected against a non-uniform large-scale background. Since emission in the 30--70~GHz range may arise from several mechanisms, including free--free, synchrotron, and anomalous microwave emission \citep{Planck2016XXV}, the \textit{Planck} morphology alone cannot be used to infer the distribution of the ambient gas or to establish a physical association with the infrared arc.

Nevertheless, the large-scale complexity seen by \textit{Planck}, together with the broader and spatially structured far-infrared emission revealed by \textit{Herschel}, provides an important context for the unusual W4 morphology. In such a non-uniform environment the shape and orientation of a wind--ISM interaction region need not be determined solely by the stellar proper-motion vector. Interaction with pre-existing interstellar structure, local gas motions, or density gradients could modify the position and curvature of the infrared ridge even when the stellar peculiar velocity is small.

Taken together, the low stellar transverse velocities, the small temperature and column-density contrasts measured with \textit{Herschel}, and the complex large-scale environment seen by \textit{Planck} argue against interpreting the W4 feature simply as an isolated high-velocity wind bow shock. An interaction between the stellar system and a structured ambient medium provides a more consistent interpretation of the observed multiwavelength morphology.

\subsection{Which stellar component is associated with the arc?}
\label{sec:discussion_components}

The identification of a single stellar component responsible for the infrared structure is not unambiguous. Different observational properties favour different components of the system.

On purely projected spatial grounds, HD~108248 and HD~108249 are the closest stars to the W4 ridge. Their mean projected star--ridge separation is approximately 0.13~pc, compared with 0.17~pc for HD~108250. In addition, HD~108248 and HD~108249 are earlier-type B stars than HD~108250 and are consequently expected to provide the stronger stellar-wind input than HD~108250 \citep{Kobulnicky2019}. 

The kinematic evidence, however, points in a somewhat different direction. The representative motion of HD~108248/HD~108249 is consistent with that of LCC, whereas HD~108250 shows a small but statistically significant peculiar proper motion. Its proper-motion vector lies relatively close to the geometrical symmetry axis of the arc, but points in the opposite sense from the infrared structure. The 28-point constant-radius test likewise shows a smaller fractional radial dispersion about HD~108250 than about the HD~108248/HD~108249 pair. This result should be treated cautiously, however, because the W4 ridge samples only a partial arc, so the apparent radial symmetry of the observed segment does not uniquely identify its physical centre.

These apparently conflicting indicators make it difficult to assign the arc uniquely to one component. The proximity, earlier spectral types, and stronger X-ray emission favour an important role for HD~108248/HD~108249, whereas the overall curvature of the traced ridge is geometrically more consistent with HD~108250 as the centre. It is also possible that the morphology is not generated by a single star. 

The complex large-scale environment discussed in Sect.~\ref{sec:discussion_arc} provides an additional reason to be cautious when identifying the responsible stellar component from the apparent curvature of the W4 ridge alone, since local ISM structure and gas motions may displace or distort the interaction region relative to the driving star.

The X-ray properties provide an additional distinction between the stellar components. The HD~108248/HD~108249 pair is associated with an X-ray luminosity of $\sim1.8\times10^{31}$~erg~s$^{-1}$, whereas HD~108250 has $L_{\rm X}\sim3\times10^{29}$~erg~s$^{-1}$. X-ray emission is well established among O and early-B stars and is generally attributed to shocks embedded in their radiatively driven winds. The empirical relation between X-ray and bolometric luminosity extends approximately to spectral types B1--B1.5 \citep{Berghoefer1997}, making the much stronger X-ray emission associated with the B0.5~IV/B1~V pair qualitatively consistent with their earlier spectral types.

The interpretation of the weaker detection associated with the B4~V star is less straightforward. Later B-type stars possess much weaker winds, and X-ray detections in this spectral range have often been discussed in terms of unresolved magnetically active lower-mass companions \citep{Berghoefer1997,Stelzer2006}. The present \textit{eROSITA} data cannot distinguish between such possibilities, and we therefore do not use the X-ray luminosity of HD~108250 as evidence for a direct interaction with the infrared arc.

More generally, X-ray observations of well-established stellar bow shocks have rarely detected diffuse emission coincident with the infrared structure. XMM--Newton observations of several runaway-star bow shocks found stellar X-ray emission but no corresponding diffuse bow-shock component \citep{Toala2017}, while the stacking analysis of \citet{Binder2019} placed an upper limit of $\lesssim2\times10^{29}$~erg~s$^{-1}$ on the average X-ray luminosity of infrared-selected Galactic bow shocks. The \textit{eROSITA} detections associated with the stellar components should therefore be interpreted primarily as stellar diagnostics rather than as evidence for X-ray emission from the extended W4 structure.

We therefore regard the W4 feature as an infrared arc associated with the environment of the $\alpha$~Crucis system, while avoiding an unambiguous classification as a bow shock driven by any single component. Higher-resolution spectroscopic observations of the surrounding gas, particularly measurements of its radial velocity and internal kinematics, would be required to establish the relative motion between the stars and the local ISM and to determine the physical origin of the structure.

\section{Conclusions}
\label{sec:conclusions}
We investigated the infrared environment and kinematics of the $\alpha$~Crucis stellar system using \textit{WISE}, \textit{Herschel}, \textit{Planck}, astrometric, and \textit{eROSITA} data. The main results can be summarized as follows:
\begin{enumerate}
\item
The \textit{WISE} W4 image reveals a prominent curved infrared ridge adjacent to the stellar system. The traced structure extends over approximately 21.1\,arcmin, corresponding to a projected length of about $0.65$~pc at the adopted distance of 106.4~pc. The median PSF-corrected width of the bright ridge is approximately 71\,arcsec, or $0.037$~pc.
\item
The minimum projected separations from the ridge are approximately $0.133$~pc for the close HD~108248/HD~108249 pair and $0.174$~pc for HD~108250. Thus, the close pair is spatially nearer to the bright infrared structure.
\item
The representative proper motion of HD~108248/HD~108249 is consistent with the mean motion of the LCC subgroup. By contrast, HD~108250 shows a modest but statistically significant peculiar proper motion, corresponding to a projected peculiar velocity of only $\sim1.8$~km\,s$^{-1}$. Its proper-motion vector lies within $\sim23^\circ$ of the geometrical symmetry axis, but points in the opposite sense from the infrared arc.
\item The \textit{Herschel}-derived maps show only small differences between the W4-defined region and the nearby background. The median dust temperature is higher by approximately 0.8~K, while the molecular hydrogen column density is higher by about $7\times10^{20}$~cm$^{-2}$. These differences are treated as descriptive contrasts rather than as statistically significant enhancements.
\item
Both HD~108248/HD~108249 and HD~108250 are detected by \textit{eROSITA}. The close pair is associated with a substantially brighter X-ray source, with $L_{\rm X}\sim$1.8$\times10^{31}$~erg~s$^{-1}$, whereas HD~108250 has $L_{\rm X}\sim3\times10^{29}$~erg~s$^{-1}$. The detected X-ray emission is therefore more plausibly associated with the stellar sources than with the extended W4 structure.
\item 
The combined evidence does not support an unambiguous interpretation of the W4 feature as a classical high-velocity stellar-wind bow shock. Instead, the arc is more likely to trace a complex interaction between the multiple stellar system and a structured ambient medium, and its association with a single stellar component cannot be established unambiguously.
\end{enumerate}

Higher-resolution spectroscopic observations of the surrounding gas would be particularly valuable for measuring its radial velocity and internal kinematics, and for determining the relative motion between the stars and the local ISM.

\backmatter

\bmhead{Acknowledgements}
We thank the anonymous referee for their helpful report that improved clarity of this work. This work was supported by the Science Committee of the Republic of Armenia through research project No.~21AG-1C044. This study has made use of data from the European Space Agency (ESA) mission \textit{Gaia}, processed by the \textit{Gaia} Data Processing and
Analysis Consortium (DPAC). This research has also made use of data products from the \textit{Herschel} Space Observatory, an ESA space observatory.

\bibliography{sn-bibliography}

\end{document}